\documentstyle[12pt]{article}

\def\beeq{\begin{eqnarray}} \def\eeeq{\end{eqnarray}}
\newcommand\mysection{\setcounter{equation}{0}\section}
\renewcommand{\theequation}{\thesection.\arabic{equation}}
\newcounter{hran} \renewcommand{\thehran}{\thesection.\arabic{hran}}

\def\bmini{\setcounter{hran}{\value{equation}}
  \refstepcounter{hran}\setcounter{equation}{0}
  \renewcommand{\theequation}{\thehran\alph{equation}}\begin{eqnarray}}

\def\bminiG#1{\setcounter{hran}{\value{equation}}
\refstepcounter{hran}\setcounter{equation}{-1}
\renewcommand{\theequation}{\thehran\alph{equation}}
\refstepcounter{equation}\label{#1}\begin{eqnarray}}

\def\emini{\end{eqnarray}\relax\setcounter{equation}{\value{hran}}\renewcommand{\theequation}{\thesection.\arabic{equation}}}

\def\ben{\begin{enumerate}}  \def\een{\end{enumerate}}
\def\bit{\begin{itemize}}    \def\eit{\end{itemize}}
\def\beq{\begin{equation}}   \def\eeq{\end{equation}}
\def\bea{\begin{eqnarray}}  \def\eea{\end{eqnarray}}
\def\nn{\nonumber}
\def\noi{\noindent}

\def\lsim{\raise0.3ex\hbox{$<$\kern-0.75em\raise-1.1ex\hbox{$\sim$}}}
\def\gsim{\raise0.3ex\hbox{$>$\kern-0.75em\raise-1.1ex\hbox{$\sim$}}}
 \def\cite#1{[\ref{#1}]}
 \def\citd#1#2{[\ref{#1},\ref{#2}]}
 
 \def\citm#1#2{[\ref{#1}--\ref{#2}]}
 \def\citr#1#2#3#4{[\ref{#1},\ref{#2},\ref{#3}--\ref{#4}]}

\begin{document}
\vbox to 1 truecm {}
\begin{center}
{\bf CONSTRAINTS ON A BRANE-WORLD FROM THE VANISHING OF THE COSMOLOGICAL CONSTANT}
\\ \vspace{1 truecm}
{\bf Ulrich Ellwanger}\footnote{email : ellwange@qcd.th.u-psud.fr}\\
Laboratoire de Physique Th\'eorique\footnote{Unit\'e Mixte de Recherche UMR 8627 - CNRS
}\\    Universit\'e de Paris XI, B\^atiment 210, 91405 Orsay Cedex,
France  \end{center} \vspace{2 truecm} 
\begin{abstract}
We derive the analogue of the vanishing of the cos\-mo\-lo\-gi\-cal 
constant in $3 + 1$ dimensions, $T_0^{\ 0} = 0$, in terms of an
integral over components of the energy-momentum tensor of a $4 + 1$
dimensional universe with parallel three-branes, and an additional
constraint local to the branes. The basic ingredients
are the existence of a static solution of the Einstein equations, and
the compactness of the 5th dimension. The corresponding constraints are
applied to a general action of scalar fields with arbitrary potentials
in the bulk and  on the branes. The equations of motion are solved in a
linearized  approximation in the 5th dimension, whereupon they require
the search for extrema of an ``effective potential'', which depends
nonlinearly on the action in the bulk and on the branes. The previous
constraints then turn into the vanishing of this ``effective
potential'' at the extremum. \end{abstract}

\vspace{2 truecm} 

\noi LPT Orsay 99-66 \\ 
\noi September 1999 \\
 
\newpage
\pagestyle{plain}
\mysection{Introduction}
\hspace*{\parindent} One of the great puzzles of fundamental physics is the smallness
of the cos\-mo\-lo\-gi\-cal constant compared to particle physics scales. From the Einstein
equations one finds that the cosmological constant can be interpreted as the
expectation value of the energy moment tensor $T_{\mu \nu}$ (which can be chosen to
be proportional to the Minkowski metric $\eta_{\mu \nu}$, given the observed
homogeneity and isotropy of the universe). From the observed values of the Hubble
parameter (or the time derivative of the Robertson-Walker scale factor), which is tiny
compared to particle physics scales, one finds that, in order to describe our present
universe, one has to require

\bminiG{1.1}
\label{1.1a}
T_i^{\ j} = \left ( T_0^{\ 0} \right ) \delta_i^{\ j} \quad ,
\eeeq  
\beeq
 \label{1.1b}
T_0^{\ 0} = 0 \qquad \hbox{(in $3 + 1$ dimensions)}
 \emini

\noi to extreme accuracy. Here $i,j = 1,2,3$ denote the 3 spatial components. In
classical field theory the expectation values of $T_{\mu \nu}$ have to be constructed
from the action and the vacuum solutions of the fields, in quantum field theory one has
to consider the quantum effective action (or to use quantum mechanical expectation
values). Whereas the first constraint (\ref{1.1a}) is easy to satisfy, provided only
scalar fields have (homogeneous) vacuum expectation values, the second constraint
(\ref{1.1b}) requires the vanishing of the scalar potential at the minimum, which
corresponds to the puzzle mentioned above. \par

Motivated, to a large extent, by the $M$-theory scenario of the strongly coupled $E_8
\times E_8$ heterotic string \citd{1r}{2r}, the cosmology of a universe with
three-branes in extra dimensions has recently been the subject of many investigations
\citm{3r}{6r}. Within $M$-theory one finds that, below the scale of Calabi-Yau
compactification, our universe is effectively five dimensional 
[\ref{1r},\ref{2r},\ref{7r}--\ref{9r}].
Observable and hidden matter lives on three-branes ($3 + 1$ dimensional space-times),
whereas gravity, moduli fields and fields originating from a 3-form in 11 dimensions
live in the $4 + 1$ dimensional bulk, whose 5th dimension is compact. \par

In the present paper we will derive the analogue of the constraints (\ref{1.1}), which
are required for a quasi-static universe, for a brane-world. The next Section 2 is kept
very general: We will just assume that our universe consists of parallel three-branes
in a $4 + 1$ dimensional bulk, and allow for an arbitrary dependence of the fields in
the bulk (and hence of the energy momentum tensor) and the components of the metric on
the extra compact 5th dimension. The requirement of a static Lorentz invariant (in $3 +
1$ dimensions) solution of the Einstein equations then leads to a new constraint on the
energy-momentum tensor in the vacuum, which differs from the constraints (\ref{1.1}).
\par

In Section 3 we will be somewhat more specific and consider the scenario with
2 three-branes at a distance $\pi \rho$, and vacuum configurations of scalar fields with
arbitrary non-linear sigma model metric and potentials in the bulk and on the branes. In
order to solve the equations of motion we resort to a small $\rho$ limit (specified
below), which allows to expand the fields in powers of the 5th dimension (up to
singularities on the branes). Then we are able to turn the constraint on the
energy-momentum tensor in the vacuum into a constraint on the action, 
which is derived in section 3. \par

In Sections 2 and 3 we check our formulas agains various published scenarios of static
universes. However, our resulting constraint on the energy-momentum tensor is very
general and applies to all particle physics phenomena as, e.g., scenarios for
supersymmetry breaking by gaugino condensation on a hidden brane. 

Finally, a brief summary is given in Section 4.

\mysection{Constraint on the energy-momentum tensor}
\hspace*{\parindent} Throughout this paper we consider a ($4 + 1$) dimensional universe
with coordinates indexed by (0, 1, 2, 3, 5). The zeroth component corresponds to the
time, and the 5th coordinate $x_5$ will often be denoted by $y$. Parallel
three-branes will be located at fixed values of $y$ which we denote by
$\widehat{y}^{(n)}$, where $n$ indexes the different branes. We are interested in
static vacuum configurations of fields and the metric which respect ($3 + 1$)
dimensional Lorentz invariance, hence the fields and the metric can only depend on the
5th coordinate $y$. For the five-dimensional metric we can then choose
 
\beq
\label{2.1}
ds^2 = a^2(y) \left ( - dx_0^2 + dx_i \ dx_j \ \delta^{ij} \right ) + b^2(y) \ dy^2
\quad .
 \eeq

\noi Below it will be convenient to write the independent components of the metric as

\beq
\label{2.2}
a(y) = e^{\alpha (y)} \quad , \qquad b(y) = e^{\beta (y)} \quad .
\eeq

The energy-momentum tensor can be chosen to be diagonal, with non-vanishing components

\beq
\label{2.3}
T_0^{\ 0} \ , \quad T_i^{\ j} = \left ( T_0^{\ 0} \right ) \delta_i^{\ j} \ , 
\quad T_5^{\ 5} \quad .
\eeq

\noi Below we will decompose the matter action into a part living in the ($4 + 1$)
dimensional bulk, and a part living on the $n$ three-branes at $\widehat{y}^{(n)}$.
Consequently $T_0^{\ 0}$ can also be decomposed into bulk and brane parts, 
whereas $T_5^{\ 5}$
depends only on the action in the bulk:

\bminiG{2.4}
\label{2.4a}
T_0^{\ 0} = T^{(bulk) \ \ 0}_{\quad \quad \ 0} + b^{-1}(y) \sum_n
\delta \left   (y - \widehat{y}^{(n)} \right ) \ T^{(brane \ n) \ \
0}_{\qquad \quad \ \ 0} \quad , 
\eeeq  \beeq
 \label{2.4b}
T_5^{\ 5} = T^{(bulk)\ \ 5}_{\quad \quad \ 5} \quad .
\emini

Now we consider the Einstein equations of a ($4 + 1$) dimensional universe. In the more
general non-static case these have been derived by Bin\'etruy, Deffayet and Langlois
\cite{4r}; in the static case they simplify to (using the notation (\ref{2.2}))

\bminiG{2.5}
\label{2.5a}
e^{-\beta} \left ( \alpha '' + 2 \alpha '^2 - \alpha ' \beta ' \right ) = {\kappa^2 \over
3} \ e^{\beta} \ T_0^{\ 0} \quad , \eeeq  \beeq
 \label{2.5b}
e^{-\beta} \ \alpha '^2 = {\kappa^2 \over 6} \ e^{\beta} \ T_5^{\ 5} \quad .
\emini

Here primes denote derivatives with respect to $y$, and $\kappa$ denotes the five
dimensional gravitational constant. The $(i,j)$ components of the Einstein equations do
not provide any independent informations. Now, from the difference of eqs. (\ref{2.5a})
and (\ref{2.5b}) one obtains

\beq
\label{2.6}
e^{-\beta} \left ( \alpha '' + \alpha '^2 - \alpha ' \beta ' \right ) = {\kappa^2 \over
3} \ e^{\beta} \left ( T_0^{\ 0} - {1 \over 2} T_5^{\ 5} \right ) \quad , \eeq

\noi which can be written as

\beq
\label{2.7}
\partial_y \left ( e^{\alpha - \beta} \alpha ' \right ) = {\kappa^2 \over 3} \ e^{\alpha
+ \beta} \left ( T_0^{\ 0} - {1 \over 2} \ T_5^{\ 5} \right ) \quad . \eeq

At this point the compactness of the 5th dimension becomes crucial. We assume that $y$
is confined into a finite interval $I$, or, alternatively, that all physical fields as
$\alpha (y)$ and $\beta (y)$ are periodic, $\alpha (y + I) = \alpha (y)$ and $\beta (y +
I) = \beta (y)$. Consequently the integral $dy$ over the interval $I$ of the left
hand-side of (\ref{2.7}) vanishes; note that this argument holds also in the presence
of branes, on which the second $y$-derivatives of $\alpha$ or $\beta$ can be singular.
Returning to the notation $a$ and $b$ instead of $\alpha$ and $\beta$, and omitting
constant factors, one derives from eq. (\ref{2.7})

\beq
\label{2.8}
\int_I dy \ a(y) \ b(y) \left ( 2 T_0^{\ 0} - T_5^{\ 5} \right ) = 0 \quad .
\eeq

Although a topological argument has been used in order to derive eq. (\ref{2.8}), it
does not represent a general topological constraint: it represents instead a necessary
condition for the existence of a static solution of the Einstein equations, i.e. a
(practically) vanishing Hubble constant. In this sense eq. (\ref{2.8}) replaces the four
dimensional constraint (\ref{1.1b}), the vanishing of the cosmological constant. \par

If one decomposes the energy-momentum tensor into its bulk and brane parts as in eqs.
(\ref{2.4}), eq. (\ref{2.8}) becomes

\beq
\label{2.9}
\int_I dy \ a(y) \left \{ b(y) \ \left ( 2T^{(bulk) \ \ 0}_{\quad \quad \ 0} - 
T^{(bulk) \ \ 5}_{\quad \quad \ 5} \right ) + 2 \sum_n \delta \left 
( y - \widehat{y}^{(n)} \right )
T^{(brane \ n )\ \ 0}_{\quad \qquad \ \ 0} \right \} = 0 \quad . \eeq

\noi Trivially, in the absence of a matter action in the bulk and for a
single three-brane, eq. (\ref{2.9}) collapses to the constraint (\ref{1.1b}). \par

Whereas the constraints (\ref{2.8}) and (\ref{2.9}) are global in the
5th dimension, it is actually possible -- under one additional
assumption -- to derive another constraint which involves just the
energy-momentum tensor on a given brane: Let us assume that a given
brane is situated at $y = 0$, and let us consider the Einstein equations
(\ref{2.5}) near $y = 0$. The right hand side of eq. (\ref{2.5a})
involves a singular term $\sim \delta(y) T^{(brane) \ \ 0}_
{ \qquad \ \ 0}$, which has to be cancelled by a singularity in
$\alpha ''$ on the left hand side. Under the assumption that $\alpha(y)$
is symmetric in $y$ (e.g. in the case of an $S^1/Z_2$ orbifold geometry,
see the next section) this condition allows to obtain $\alpha '$ on the
brane: Now $\alpha$ can only depend on $|y|$, and one has

\beq
\label{2.10}
\alpha ''(y) = 2 \alpha '(0) \delta(y) + \dots
\eeq

\noi near $y=0$, where the dots denote regular terms. Then, one obtains
from eq. (\ref{2.5a})

\beq
\label{2.11}
\alpha '(0) = {\kappa^2 \over 6} e^{\beta}T^{(brane) \ \ 0}_{ 
\qquad \ \ 0}
\eeq

\noi and hence, from eq. (2.5b) at $y=0$,

\beq
\label{2.12}
{\kappa^2 \over 6} (T^{(brane) \ \ 0}_{\qquad \ \ 0})^2
- T^{(bulk) \ \ 5}_{\quad \quad \ 5}(0) = 0 \quad .
\eeq

This condition can also be derived from the results of Bin\'etruy,
Deffayet and Langlois [4], and coincides with one of the constraints
required for the solution of Randall and Sundrum [5]. It should be
noted, however, that -- up to the symmetry assumption on
$\alpha(y)$ --
it is completely independent from the form of the energy-momentum tensor
along the 5th dimension. \par

Next, we would like to express the general constraint (\ref{2.8}) or 
(\ref{2.9}) in terms of an action involving matter in the bulk and on 
the branes. Since the only possible Lorentz
invariant vacuum configurations are $y$-dependent scalar fields $\varphi^i
(y)$ (up to duality transformations to 3-forms in 5 dimensions), we will restrict
ourselves to a general non-linear sigma model action (neglecting higher derivatives) with
sigma model metric ${\cal G}_{ij}(\varphi )$ and arbitrary potentials in the bulk and
on the branes. (Actually, in the presence of gauge fields in the bulk, we could also
have Wilson lines along the compact 5th dimension. They do, however, not contribute
to the energy-momentum tensor). \par

Taking already into account that, in the vacuum, all derivatives of the fields
$\varphi^i$ except with respect to $y$ vanish, the general action in the bulk and on the
branes reads (using the metric (\ref{2.1}))

\bea
\label{2.13}
S(\varphi ) &=& - \int d^5 x \ a^4(y) \Big \{ b(y) \left ( 
{\cal G}_{ij} (\varphi ) \ \partial_y \varphi^i\ \partial^y \varphi^j 
+ V^{(bulk)}(\varphi ) \right ) \nn \\
&& + \sum_n
\delta \left ( y - \widehat{y}^{(n)} \right ) V^{(brane \ n)} (\varphi ) \Big \} \quad .
\eea

\noi The $(0, 0)$ and $(5,5)$ components of the energy-momentum tensor 
can easily be derived from eq. (\ref{2.13}):

\bminiG{2.14}
\label{2.14a}
T_0^{\ 0} &=& - \Big \{ b^{-2}(y) {\cal G}_{ij} (\varphi ) \ 
\partial_y \varphi^i\ \partial_y \varphi^j 
+ V^{(bulk)}(\varphi ) \nn \\ 
&&
+ b^{-1}(y)\sum_n
\delta \left ( y - \widehat{y}^{(n)} \right ) 
V^{(brane \ n)} (\varphi ) \Big \} \quad ,
\eeeq  
\beeq
\label{2.14b}
T_5^{\ 5} = b^{-2}(y) {\cal G}_{ij} (\varphi ) \ 
\partial_y \varphi^i\ \partial_y \varphi^j 
- V^{(bulk)}(\varphi )
\emini

\noi Inserting eqs. (2.14) into eqs. (\ref{2.8}) or (\ref{2.9}), these
constraints become

\bea
\label{2.15}
&&\int_I dy \ a(y) \Big \{ 3b^{-1}(y) \ 
{\cal G}_{ij} (\varphi ) \ \partial_y \varphi^i\ \partial_y \varphi^j 
+ b(y) V^{(bulk)}(\varphi ) \nn \\
&&\qquad \qquad + 2 \sum_n \delta \left ( y -
\widehat{y}^{(n)} \right ) V^{(brane \ n)} (\varphi ) \Big \} = 0 \eea 

\noi where the fields $\varphi^i(y)$ (as well as $a(y)$, $b(y)$) are solutions of the
equations of motion. \par

In the form of eq. (2.15), the general condition for a static 
universe can again be
compared to the constraints in more specific scenarios as the 
non-conventional 
cosmology from a brane universe in \cite{4r},
and the generation of hierarchies in \cite{5r}. 
In both cases no scalar fields
$\varphi^i$ exist, i.e. the first term in (2.15) vanishes, and the potentials are
just constants (with $V^{(bulk)} = 0$ in \cite{4r}). Inserting the solution for $a(y)$
(and $b(y) = 1$) of \cite{5r} into (\ref{2.15}), and using $V^{(brane \ 2)} = -
V^{(brane \ 1)}$, one obtains indeed a second 
constraint on the vacuum energies in the
bulk and on the branes, as compared to eq. (\ref{2.12}), 
as in \cite{5r}. Likewise, for $V^{(bulk)} = 0$ only the last
term survives in (\ref{2.15}), and the resulting constraint agrees with eq. (46) in
\cite{4r}. (Notably, a static universe had not been assumed in \cite{4r} in order to
obtain this constraint; there it follows generally from the solution for $a(y)$ in the
presence of 2 three-branes without an energy-momentum tensor in the bulk). \par

It should be noted that, in general, the local constraints (2.12) on
each brane and the global constraint (2.9) are independent, which
explains the presence of two conditions for a static universe in
\cite{4r} and \cite{5r}. \par

Whereas our previous results (\ref{2.8}), (\ref{2.9}), (\ref{2.12}) 
and (\ref{2.15}) are very general,
their practical application -- at 
least in the presence of matter in the bulk -- is
complicated by the fact that it requires the knowledge of the vacuum solutions for
$a(y)$, $b(y)$ and $\varphi^i (y)$. In the ($3 + 1$) dimensional case, the analogous
constraint (\ref{1.1b}) implies the well-known condition of a vanishing scalar potential
at its minimum. Likewise, we would like to translate our previous results into a
condition on the action (\ref{2.13}). In the case of 2 three-branes which are
sufficiently close to each other, we are able to derive such a condition in the next
chapter.

\mysection{2 close three-branes} 
\hspace*{\parindent} Motivated by the compactification of $M$-theory from 11 to 5
dimensions on a Calabi-Yau manifold $X_{CY}$ \citr{1r}{2r}{7r}{9r}, we will now 
consider a
5th dimension with the geo\-me\-try of an orbifold $S^1/Z_2$. We choose $y$ in the
interval $y \subset [- \pi \rho , \pi \rho]$ with the endpoints being identified. Two
three-branes are located at $\widehat{y}^{(1)} = 0$ and $\widehat{y}^{(2)} = \pi \rho$,
and the $Z_2$ orbifold symmetry acts as $y \to - y$. \par

In this scenario the $h^{1,1}$ moduli of $X_{CY}$ appear in 5 dimensions as scalar
fields $\varphi ^i$ with non-trivial potentials both in the bulk and on the branes
\citm{7r}{9r}. Notably, these potentials are proportional to coefficients $\alpha_i$,
which are given by integrals of $Tr \ R \wedge R$ over corresponding four-cycles $C_i$
in $X_{CY}$ \cite{8r}. In general we have

\beq
\label{3.1}
\alpha_i \sim \alpha \quad \hbox{with} \qquad \alpha \sim {\cal O}\left (
M_{GUT}^4/M_{11}^3 \right ) \quad ,  \eeq 

\noi where $M_{GUT}^6$ is the inverse size of $X_{CY}$, and $M_{11}$ the scale of the
eleven-dimensional gravitational coupling. In terms of $\alpha$ the potentials of the
moduli $\varphi^i$ are of the respective orders \cite{8r}
 
\bminiG{3.2}
\label{3.2a}
V^{(brane \ 1)} (\varphi^i) = - V^{(brane \ 2)} (\varphi^i) = {\cal O}(\alpha ) \quad ,
\eeeq  \beeq
 \label{3.2b}
V^{(bulk)}(\varphi^i) = {\cal O}(\alpha^2) \quad .
\emini

\noi Comparing different terms in the equations of motion for the fields $\varphi^i
(y)$ (see below) one then finds that an expansion of $\varphi^i (y)$ in powers of $y$
converges for

\beq
\label{3.3}
\alpha \rho \ll 1 \quad .
\eeq  

A phenomenological motivation for the inequality (\ref{3.3}) within the $M$ theory
scenario can be derived from the relation obtained in \cite{2r} between the gauge
couplings on the hidden and observable branes, respectively. Given a small value of
$\alpha_{GUT}$, (\ref{3.3}) follows from requiring a small gauge coupling of the $E_8$
gauge symmetry on the hidden brane at the scale $M_{11}$, such that gaugino condensation
occurs only at a scale much below $M_{11}$ and a susy breaking scale far below
$M_{Planck}$ is generated \cite{7r}. On the other hand, the inequality 
(\ref{3.3}) is certainly not
strong given the measured value of Newton's constant and the preferred values of
$\alpha_{GUT}$ and $M_{GUT}$ \citd{2r}{7r}. \par

Independently from this possible (and possibly doubtful)
phenomenological motivation we will now show how, to lowest non-trivial
order in $\alpha \rho$, the constraints  derived in Section 2 turns into
a constraint on the action. To this end we expand the  fields
$\varphi^i$ and the components of the metric in powers of $y$ (only
$|y|$ appears, once one requires continuity across the second brane, or
the fields to be even under the $Z_2$ orbifold symmetry): 
 
\bminiG{3.4}
\label{3.4a}
\varphi ^i (y) = \varphi^i_0 + \alpha \varphi^i_1 |y| + {\cal O}(\alpha^2)
\eeeq  \beeq
 \label{3.4b}
a(y) = a_0 \left ( 1 + \alpha \ a_1 |y| + {\cal O}(\alpha^2) \right )
\eeeq 
\beeq
\label{3.4c}
b(y) = 1 + \alpha \ b_1|y| + {\cal O}(\alpha^2) \quad .
  \emini

\noi In (\ref{3.4c}) we have used the freedom in the definition of $y$ such that $b(0) =
1$. Below we will need the second $y$ derivatives of $\varphi ^i(y)$ and $a(y)$, which are
singular on the branes:

\bminiG{3.5}
\label{3.5a}
{\varphi^i} ''(y) = 2 \alpha \ \varphi^i_1 \left ( \delta (y) - \delta (y - \pi \rho )
\right ) \quad ,  \eeeq  \beeq
 \label{3.5b}
a''(y) = 2 \alpha \ a_0 \ a_1 \left ( \delta (y) - \delta (y - \pi \rho ) \right )
\quad .
  \emini

\noi Next we consider the equations of motion of the fields $\varphi^i$, as derived from
the general action (\ref{2.13}):

\bea
\label{3.6}
&&{1 \over{b(y)}} \ {\cal G}_{ij,k} \ \partial_y \varphi^i \ 
\partial_y \varphi^j
- {2 \over a^4(y)} \ \partial_y \left ( {a^4(y) \over b(y)} \ 
{\cal G}_{ik} \ \partial_y
\varphi^i \right ) \nn \\
&&+ b(y) \ V^{(bulk)}_{\quad \quad \ ,k} + \sum_n \delta \left ( y - \widehat{y}^{(n)}
\right ) V^{(brane \ n)}_{\quad \qquad \ ,k} = 0 \quad . \eea

\noi Whereas the terms in the bulk of ${\cal O}(\alpha^2)$ in (\ref{3.6}) determine the
terms in $\varphi^i (y)$ of quadratic and higher order in $y$ neglected in (\ref{3.4a}),
the singular terms on the branes of ${\cal O}(\alpha)$ --including contributions from
the second derivatives (\ref{3.5a}) of $\varphi^i$-- determine the coefficients
$\varphi_1^i$. Using

\beq
\label{3.7}
V^{(brane \ 1)} = - V^{(brane \ 2)} \equiv V^{(br)}
\eeq

\noi and (\ref{3.5a}), the singular terms in (\ref{3.6}) give

\beq
\label{3.8}
\varphi_1^i = {1 \over 4 \alpha} \ {\cal G}^{ik}(\varphi_0) \ V^{(br)}(\varphi_0)_{,k}
\eeq

\noi with ${\cal G}^{ik} {\cal G}_{kj} = \delta^i_{\ j}$. (Here, as well as in the case of
eq. (\ref{3.10}) below, eq. (\ref{3.7}) is required for the consistency of the ansatz
(\ref{3.4})). Hence we have $\partial_y \varphi^i \sim {\cal O}(\alpha )$, and the first
two terms in eq. (\ref{2.15}) are both of ${\cal O}(\alpha^2)$. Given the orbifold
geometry and eq. (\ref{3.7}), the last term in eq. (\ref{2.15}) reads

\beq
\label{3.9}
2a(0) \ V^{(br)} ( \varphi (0) ) - 2a (\pi \rho ) \ V^{(br)} \ (\varphi (\pi \rho ))
\quad . \eeq

\noi In order to evaluate $a(\pi \rho )$ to ${\cal O}(\alpha )$ 
we need $a_1$ in eq. (\ref{3.4b}). As in the
case of $\varphi_1^i$, this coefficient is obtained from the singular part of the
corresponding equation of motion, i.e. the Einstein equation (\ref{2.5a}) with 
$T_0^{\ 0}$
as in eq. (\ref{2.4a}). To leading order in $\alpha$ one finds from (\ref{2.5a})

\beq
\label{3.10}
a_1 = -{\kappa^2 \over 6 \alpha} \ V^{(br)}(\varphi_0) \quad .
\eeq

\noi Finally we can use, in the last term in (\ref{3.9}),

\bea
\label{3.11}
V^{(br)} (\varphi (\pi \rho )) &=& V^{(br)} (\varphi (0)) + \pi \rho \alpha \ \varphi_1^i
\ V^{(br)}(\varphi(0))_{,i} + {\cal O}(\alpha^3) \nn \\
&=& V^{(br)}(\varphi_0) + {\pi \rho \over 4} \ V^{(br)}(\varphi_0)_{,k} \ {\cal
G}^{ki}(\varphi_0) \ V^{(br)}(\varphi_0)_{,i} + {\cal O}(\alpha^3) .
 \eea

\noi Terms of ${\cal O}(\alpha )$ cancel in (\ref{3.9}) and hence in eq. (\ref{2.15}),
and the terms of ${\cal O}(\alpha^2)$ in eq. (\ref{2.15}) read altogether

\bea
\label{3.12}
&&\int_I dy \ a_0 \left \{ {3 \over 16} V^{(br)}(\varphi_0)_{,i} \ 
{\cal G}^{ij}(\varphi_0) V^{(br)}(\varphi_0)_{,j} +
V^{(bulk)}(\varphi_0) \right \} \nn \\
&&+ 2a_0 \ \pi \rho \left \{ {\kappa^2 \over 6} \left (
V^{(br)}(\varphi_0)\right )^2 - {1 \over 4} V^{(br)}(\varphi_0)_{,i} \
{\cal G}^{ij}(\varphi_0) V^{(br)}(\varphi_0)_{,j} \right \} = 0 \ .
\eea

\noi Using the size $2\pi \rho$ of the interval $I$, the former constraint (\ref{2.15})
becomes after dividing by $\pi \rho a_0$:

\beq
\label{3.13}
- {1 \over 8} V^{(br)}(\varphi_0)_{,i} \ {\cal G}^{ij}(\varphi_0)
V^{(br)}(\varphi_0)_{,j} + {\kappa^2 \over 3} \left ( V^{(br)}(\varphi_0)\right )^2 +
2V^{(bulk)}(\varphi_0) = 0 \quad .  \eeq

\noi Given the previous approximations, terms of relative order $\alpha \rho$ have been
neglected in eq. (\ref{3.13}). \par

Turning to the local constraint (\ref{2.12}) on the branes one observes,
using eqs. (2.14) and (3.8), that they coincide with eq. (\ref{3.13}) to
 ${\cal O}(\alpha^2)$. \par

It remains to determine the constant modes $\varphi_0^i$ from the scalar equations of
motion (\ref{3.6}). To this end we multiply eq. (\ref{3.6}) with $a^4(y)$ (which brings
it back to its original form) and integrate over the compact interval $I$, whereupon
the second term in (\ref{3.6}) vanishes. Thus eq. (\ref{3.6}) turns into

\bea
\label{3.14}
&&\int_I dy \ a^4(y) \Big \{ b^{-1}(y) \ {\cal G}_{ij,k} \ \partial_y 
\varphi^i \
\partial_y \varphi^j + b(y) V^{(bulk)}_{\quad \quad ,k} \nn \\
&& + \sum_n \delta \left (
y - \widehat{y}^{(n)} \right ) V^{(brane \ n)}_{\qquad \quad \ ,k} \Big \} = 0 \quad .
\eea  

Now we treat eq. (\ref{3.14}) as the constraint (\ref{2.15}) before, i.e. we 
insert the
ans\"atze (\ref{3.4}) for $\varphi ^i$ and $a$, use the solutions (\ref{3.8}) and
(\ref{3.10}) for $\varphi ^i_1$ and $a_1$, and keep only the leading terms of ${\cal
O}(\alpha^2)$. Then eq. (\ref{3.14}) can be brought into the form

\beq
\label{3.15}
{\delta \over \delta \varphi_0^k} \left [ - {1 \over 8} V^{(br)}_{\quad \ ,i} \ {\cal
G}^{ij}\ V^{(br)}_{\quad \ , j} + {\kappa^2 \over 3} \left ( V^{(br)} \right )^2 +
2V^{(bulk)} \right ] = 0 \eeq

\noi whose solution determines $\varphi_0^i$. Evidently the expression in parenthesis in
(\ref{3.15}), which coincides with the left hand-side of eq. (\ref{3.13}), plays the
role of an ``effective potential'': The equations of motion of the constant modes
$\varphi_0^i$ correspond to the search for its extrema, and it has to vanish at the
extremum, if the 5 dimensional analog of the 4 dimensional cosmological constant, i.e.
the left hand-sides of eqs. (\ref{2.8}), (\ref{2.9}), (\ref{2.15}) or (\ref{3.13}), are
required to vanish. However, we recall again that eqs. (\ref{3.13}) and (\ref{3.15})
have been derived using the linearized solutions (\ref{3.4}) in the bulk, and are
subject to corrections of ${\cal O}(\alpha \rho)$. (Within this approximation it is
amusing to note that the first two terms in (\ref{3.13}) and (\ref{3.15}) show some
formal similarity to the scalar potential in supergravity in $d=4$, once $V^{(br)}$ is
identified with the superpotential, although no reference to 
supersymmetry has ever
been made in its derivation. We have no explanation for this fact at present). \par

Finally we will apply our results to the sigma model action of the 
universal Calabi-Yau
modulus, using the formulation in \cite{8r} (the corresponding field has been called
$V$ there, but we will denote it by $\varphi$). An exact static solution to the Einstein
equations and the $\varphi$ equations of motion has been obtained in \cite{8r}, but of
course our approximate equations (\ref{3.13}) and (\ref{3.15}) should also apply.
Indeed, in our notation the action of the single scalar field $\varphi$ is described
by

\beq
\label{3.16}
( {\cal G}(\varphi ))^{-1} = 4 \kappa^2 \varphi^2 \quad , \qquad V^{(br)}(\varphi ) = -
{\sqrt{2} \alpha \over \kappa^2 \varphi} \quad , \qquad V^{(bulk)}(\varphi ) =
{\alpha^2 \over 6 \kappa^2 \varphi^2}  \eeq

\noi and one finds that the left hand-side of eq. (\ref{3.13}), and hence the
parenthesis in eq. (\ref{3.15}), vanish identically in $\varphi_0$. 
This results agrees with the presence of arbitrary integration constants in the
solution of \cite{8r}, some combination of which has been fixed by our convention $b(0)
= 1$. 

\mysection{Summary and Conclusions} 
\hspace*{\parindent} In the present paper we have derived the analogue of the vanishing
of the cos\-mo\-lo\-gi\-cal constant in $3 + 1$ dimensions, $T_0^{\ 0} = 0$, in terms of the
energy-momentum tensor in the vacuum of a $4 + 1$ dimensional brane-universe. The
corresponding general constraint takes the form of eqs. (\ref{2.8}) or (\ref{2.9}), and
its application requires -- as in the $3 + 1$ dimensional case -- the knowledge of the
solutions of the equations of motion. In the case of a general action of scalar fields
with arbitrary potentials in the bulk and on the branes, these solutions cannot be
obtained explicitly. They can still be constructed, however, in the linearized
approximation, where only a linear dependence of the fields and the metric in the 5th
coordinate $y$ is taken into account. \par

This ansatz can be motivated in the $M$-theory scenario, where the scalar fields in $4
+ 1$ dimensions arise as moduli of the Calabi-Yau space: 
First, one obtains the mirror
symmetry (\ref{3.7}) among the potentials on the two three-branes, which is required
for the consistency of this ansatz, and second, the potentials involve dimensionful
pa\-ra\-me\-ters $\alpha_i \sim \alpha$, which allow to define a consistent dimensionless
expansion parameter $\alpha \rho$. \par

Within this approximation we have found that the equations of motion of the cons\-tant
modes $\varphi_0^i$ of the scalar fields in the bulk correspond to the search for
extrema of an ``effective potential'', which has to be constructed in terms of the
potentials on the branes, and the potential and the non-linear sigma model metric in the
bulk. The necessary condition for a static $4 + 1$ dimensional brane-universe then
becomes the condition of a vanishing ``effective potential'' at the extremum, in analogy
to the $3 + 1$ dimensional case. Clearly we expect to find more consequences
of our general constraint derived in section 2, beyond the linearized approximation. \\

\noindent {\Large \bf Acknowledgement}\par \vskip 3 truemm
We thank P. Bin\'etruy and C. Deffayet for stimulating discussions. 

\newpage
\def\labelenumi{[\arabic{enumi}]}
\noindent
{\Large \bf References}
\ben
\item\label{1r} P. Horava, E. Witten, Nucl. Phys. {\bf B460} (1996) 506, Nucl. Phys.
{\bf B475} (1996) 94.  
\item\label{2r} E. Witten, Nucl. Phys. {\bf B471} (1996) 135. 
\item\label{3r} N. Kaloper, I. Kogan, K. Olive, Phys. Rev. {\bf D57} (1998) 7340.
Erratum ibid. {\bf D60} (1999) 049901;\\
N. Arkani-Hamed, S. Dimopoulos, G. Dvali, Phys. Rev. {\bf D59} (1999) 086004; \\
K. Benakli, Int. J. Mod. Phys. {\bf D8} (1999) 153, and Phys. Lett. {\bf
B447} (1999) 52; \\
N. Kaloper, A. Linde, Phys. Rev. {\bf D59} (1999) 101303; \\
D. Lyth, Phys. Lett. {\bf B448} (1999) 191; \\
G. Dvali, S.-H. Tye, Phys. Lett. {\bf B450} (1999) 72; \\
C. Csaki, M. Graesser, J. Terning, Phys. Lett. {\bf B456} (1999) 16; \\
G. Dvali, Phys. Lett. {\bf B459} (1999) 489; \\
A. Lukas, B. Ovrut, D. Waldram, hep-th/9806022 and hep-th/9902071; \\
N. Arkani-Hamed, S. Dimopoulos, J. March-Russell, hep-th/9809124; \\
H. Reall, hep-th/9809195; \\
C. Kolda, D. Lyth, hep-ph/9812234; \\
M. Gogberashvili, hep-ph/9812296; \\
A. Mazumdar, hep-ph/9902381; \\
N. Arkani-Hamed, S. Dimopoulos, M. Kaloper, J. March-Russell, hep-ph/9903224; \\
H. Chamblin, H. Reall, hep-th/9903225; \\
G. Dvali, M. Shifman, hep-th/9904021;\\
A. Riotto, hep-ph/9904485.
\item\label{4r} P. Bin\'etruy, C. Deffayet, D. Langlois, hep-th/9905012. 
\item\label{5r} L. Randall, R. Sundrum, hep-ph/9905221. 
\item\label{6r} N. Kaloper, hep-th/9905210; \\
E. Halyo, hep-ph/9905244; \\
T. Nihei, hep-ph/9905487; \\
L. Randall, R. Sundrum, hep-th/9906064; \\
A. Kehagias, hep-th/9906204; \\
C. Csaki, M. Graesser, C. Kolda, J. Terning, hep-ph/9906513; \\
J. Cline, C. Grojean, G. Servant, hep-ph/9906523; \\
D. Chung, K. Freese, hep-ph/9906542; \\
P. Steinhard, hep-th/9907080; \\
N. Arkani-Hamed, S. Dimopoulos, G. Dvali, N. Kaloper, hep-th/9907209; \\
W. Goldberger, M. Wise, hep-ph/9907218 and hep-ph/9907447; \\
I. Oda, Hep-th/9908104; \\
J. Lykken, L. Randall, hep-th/9908076; \\
T. Li, hep-th/9908174; \\
C. Csaki, Y. Shirman, hep-th/9908186; \\
K. Dienes, E. Dudas, T. Ghergetta, hep-ph/9908530; \\
A. Nelson, hep-th/9909001; \\
H. B. Kim, H. D. Kim, hep-th/9909053;\\
K. Behrndt, M. Cvetic, hep-th/9909058; \\
K. Skenderis, P. Townsend, hep-th/9909070; \\
H. Hatanka, M. Sakamoto, M. Tachibana, K. Takenaga, hep-th/9909076.
\item\label{7r} T. Banks, M. Dine, Nucl. Phys. {\bf B479} (1996) 173. 
\item\label{8r} A. Lukas, B. Ovrut, K. Stelle, D. Waldram, Phys. Rev. {\bf D59}
(1999) 086001 and Nucl. Phys. {\bf B552} (1999) 244.     
\item\label{9r} J. Ellis, Z. Lalak, S. Pokorski, W. Pokorski, 
Nucl. Phys. {\bf B540} (1999) 149; \\
J. Ellis, Z. Lalak, W. Pokorski, hep-th/9811133.       
\een

\end{document}